# Soft-Error Characterization and Hardening Trade-offs in Static PCHB Asynchronous Circuits

Ramya Karri, Srija Rasoori, and Ashiq A. Sakib
Dept. of Electrical and Computer Engineering
Southern Illinois University Edwardsville, Edwardsville, IL, USA
{rkarri; srasoor; asakib}@siue.edu

***Abstract*—Pre-Charge Half Buffer (PCHB) is a promising asynchronous digital design paradigm for harsh-environment operation; however, its soft-error characteristics remain largely unexplored. This paper presents a systematic soft-error characterization and hardening trade-off analysis for static PCHB circuits. A controlled transistor-level fault-injection framework is developed to extract polarity-dependent critical charge at internal nodes. Vulnerability nodes are identified based on extensive simulation. Four mitigation strategies, double-sided Schmitt trigger, single-sided Schmitt trigger, transmission-gate reinforcement, and duplication-based redundancy, are implemented and evaluated across five representative PCHB cells. Comprehensive resilience-overhead comparisons in delay, energy, and area are reported, leading to architecture-specific hardening guidelines for robust static PCHB design.**

## I. INTRODUCTION

Digital integrated circuits (ICs) designed for operation under extreme environments are essential to applications of national importance, including space exploration, deep-sea missions, and defense surveillance systems. Ensuring dependable operation under such conditions, however, remains a significant challenge. Aggressive device and supply voltage scaling increase susceptibility to radiation, electromagnetic interference (EMI), and other noise sources [1]. Radiation-induced soft errors, known as single-event upsets (SEUs), occur when high-energy particles strike sensitive nodes, generating transient current pulses that may either dissipate or get latched and propagate through the logic network, causing unintended state changes [2]. The risk of SEUs is particularly pronounced in conventional clock-driven synchronous circuits at advanced technology nodes, where reduced supply voltages and tighter timing margins amplify sensitivity to process, voltage, and temperature (PVT) variations. Such transient faults can disrupt timing relationships, cause loss of synchronization, and ultimately lead to functional failure. In contrast, quasi-delay-insensitive (QDI) asynchronous (clockless) architectures mitigate several timing-related vulnerabilities due to their insensitivity to precise delay assumptions and can offer improve tolerance to EMI and PVT variations, making them a compelling alternative for harsh-environment operation [3, 4], although they are not entirely immune to SEU effects [5].

Null Convention Logic (NCL) [6] and Pre-Charge Half Buffers (PCHB) [7] are two major QDI paradigms. Among these, NCL has historically received greater research attention. This trend is largely attributable to its synchronous-like framework, which can leverage existing computer-aided design (CAD) infrastructures with relatively minor adaptation. Consequently, prior investigations into soft-error resilience and architectural hardening in QDI systems have focused predominantly on NCL-based implementations, including node-level fault characterization, Schmitt-trigger-based filtering, and duplication-based redundancy techniques [3, 8-10].

Although NCL and PCHB have similarities, their framework, evaluation, and holding mechanisms differ substantially. For instance, static PCHB circuits employ a state-retention mechanism based on internal regenerative feedback rather than threshold-based input consensus in NCL. This structural distinction alters transient fault dynamics, node-level sensitivity, and protocol-level error behavior. As a result, resilience characteristics and mitigation assumptions established for NCL may not directly transfer to static PCHB. Despite the performance and pipeline advantages of static PCHB, its soft-error behavior, dominant vulnerability points, and hardening trade-offs have not been systematically studied. A dedicated reliability analysis is therefore necessary to identify the true critical nodes, quantify their sensitivity, and determine the effectiveness of existing hardening techniques within this distinct architectural context. This work addresses the identified gap through the following contributions:

- We develop a fault-injection and modeling framework tailored specifically to static PCHB circuits, explicitly capturing the impact of regenerative feedback-based state retention on transient fault dynamics.
- Through analytical modeling and simulation, we identify the dominant soft-error sensitive points in static PCHB that exhibit significantly lower critical charge than other nodes,
- We evaluate and compare four mitigation strategies, double-sided Schmitt trigger, single-sided Schmitt trigger, transmission gate reinforcement, and duplication-based redundancy, quantifying their impact on critical charge, area, propagation delay, and switching energy.
- Finally, we derive design recommendations for selecting appropriate hardening strategies under different reliability and cost constraints based on a hardening trade-off analysis.

To the best of our knowledge, systematic soft-error characterization and comparative hardening trade-off analysis for static PCHB circuits have not been previously reported. The remainder of the paper is organized as follows. Section II reviews the static PCHB framework. Section III presents soft-error modeling and vulnerability analysis. The different mitigation schemes are discussed in Section IV, followed by the simulation results and comparative evaluation in Section V. Section VI concludes the paper.

## II. PCHB Background

QDI PCHB circuits differ structurally and operationally from conventional synchronous/Boolean implementations in several keyways. PCHB employs dual-rail (DR) encoding rather than the single-rail representation used in Boolean logic. A DR variable, $X$, is encoded using two wires/rails, $\{X^1, X^0\} \in \{0,1\}$, enabling three valid signal states: DATA0 ($X^1 = 0, X^0 = 1$), DATA1 ($X^1 = 1, X^0 = 0$), and NULL ($X^1 = X^0 = 0$). DATA0 and DATA1 correspond to Boolean logic 0 and 1, respectively, while NULL serves as a spacer state indicating the unavailability of DATA. The condition ($X^1 = \mathrm{X}^0 = 1$) is disallowed and represents an illegal state, referred to as DATAX. Each PCHB gate inherently possesses storage capability. That is, apart from implementing a combinational logic function, a PCHB gate also behaves as a latch. Consequently, an interconnected network of purely combinational PCHB gates naturally forms a pipeline structure analogous to a synchronous pipeline. This is unlike NCL, which has registration and logic separate. Synchronization is achieved through a four-phase handshaking protocol, like other QDI paradigms. Each PCHB component includes a single request input (*Rack*) and a single acknowledge output (*Lack*) to coordinate communication and maintain an alternating sequence of DATA and NULL to distinguish between two distinct DATA wavefronts in the absence of a global clock.

PCHB gates and standard components can be implemented in semi-static as well as static manner. Fig. 1a depicts a generic semi-static 2-input PCHB function template with dual-rail inputs *A* and *B* and one dual-rail output *Z*. The *Func.* block implements the intended logic function. When both *Rack* and *Lack* are in the *request-for-data (rfd*; i.e., logic 1*)* state and inputs *A* and *B* carry valid DATA, the gate evaluates and produces DATA at the output. If one handshake signal is *rfd* while the other is *request-for-null* (*rfn*; i.e., logic 0), the output is preserved by a weak feedback inverter network. When both *Rack* and *Lack* are in the *rfn* state, the output is pre-charged to NULL. Input and output completion detection circuits (*ICD* and *OCD*), implemented using NOR-based structures, determine whether the respective signals represent DATA or NULL. Their outputs are combined through a resettable C-element to generate the acknowledgment signal, *Lack*. *Lack* transitions to *rfn* when all inputs and outputs are DATA, and returns to *rfd* once all signals return to NULL.

Fig. 1b illustrates a static PCHB function template [11] producing a single dual-rail output, $Z$. Each output rail has its own *SET* function and *Hold* function blocks, where the *Hold* logic is the logical complement of the corresponding *SET* logic. Sleep transistors are driven by a *sleep* control signal derived from $K_i$ and $K_o$ signals using a TH22n NCL gate. In this static realization, $K_i$ and $K_o$ serve roles analogous to *Rack* and *Lack* in the semi-static style. The TH22n gate operates as a two-input inverted C-element and is initialized in the reset-to-0 state. The *sleep* signal is asserted when both $K_i$ and $K_o$ are *rfn*, forcing the output *Z* to transition to NULL. Conversely, when $K_i$ and $K_o$ are both *rfd*, *sleep* is de-asserted, enabling *Z* to evaluate to DATA provided the inputs carry valid DATA. After *Z* transitions to DATA, the feedback transistors maintain that state until *sleep* is asserted again. Consequently, if *Z* holds DATA while *sleep* remains de-asserted, it will not return to NULL even if all dual-rail inputs transition to NULL, behavior consistent with the requirements of the PCHB handshaking protocol. Fig. 1c presents the sleep-generation circuitry for a 2-input, single-output PCHB circuit. The first-level TH12n gates implement NOR-based detection to determine NULL or DATA conditions at both inputs and output, similar in function to the ICD/OCD structures in semi-static designs. A TH33 gate then acts as a 3-input C-element to produce the acknowledge signal $K_o$. Finally, $K_i$ and $K_o$ feed a resettable TH22n gate that generates the *sleep* control signal. Although the static implementations require more area, they are faster and more energy efficient than the equivalent semi-static implementations and can operate correctly at very low supply voltages.

## III. Soft Error Modeling and Vulnerability Analysis of Static PCHB Standard Cells

In radiation-prone environments, high-energy neutrons or alpha particles may deposit charge at sensitive transistor nodes, producing a transient current spike. Depending on the electrical and logical state of the struck node, the resulting spike may either get filtered through electrical masking or propagate and induce unintended switching in logic elements. For static PCHB circuits, which rely on regenerative feedback for state retention, understanding node-level sensitivity is essential for reliability characterization. The particle-induced transient current was initially modeled based on the equation presented in [1]:

$$I(t) = \frac{2Q}{T\sqrt{\pi}}\sqrt{\frac{t}{T}}\,e^{-\frac{t}{T}} \;\ldots\ldots\ldots\; (1)$$

where $Q$ denotes the accumulated charge and $T$ is a technology dependent time constant that differs for NMOS and PMOS. While this exponential form captures charge collection dynamics accurately, a trapezoidal approximation was adopted for transient simulation efficiency and simplicity, consistent with established soft-error evaluation methodologies [8-9]. The trapezoidal pulse (Fig. 2a) is parameterized by peak current ($I_{\max}$), rise time ($t_{rise}$), fall time ($t_{fall}$), and pulse width ($pw$). These parameters were selected such that the total injected charge matches that of the Hazucha model for the corresponding technology node. Technology-dependent values for 350 nm, 130 nm, and 45 nm nodes were adopted from [9], and 32 nm parameters were extrapolated according to device-scaling trends and collection time constants reported in [9]. The resulting parameters are summarized in Table I.

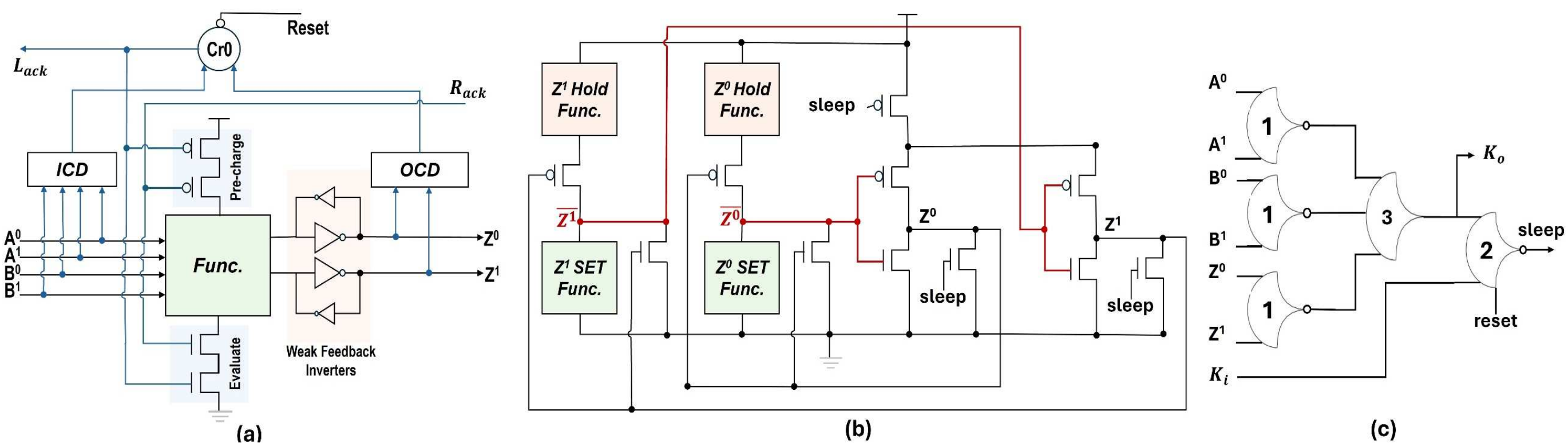

Fig. 1: PCHB design templates: (a) Semi-static, (b) static; and (c) *sleep* generation network for a 2-input single-output PCHB circuit.

For a given strike location (e.g., NMOS or PMOS devices), and prevailing logic state, four classes of output-level soft-error manifestations may occur: positive transient glitch (0→1→0), negative transient glitch (1→0→1), static positive upset (0→1), and static negative upset (1→0). Transient glitches may either self-attenuate due to electrical or logical masking or be captured by the intrinsic feedback and state-holding mechanisms of the PCHB structure. For instance, if a particle-strike perturbs a feedback-controlled node beyond its threshold, the regenerative loop may amplify and latch the disturbance, converting a transient glitch into a static upset. Therefore, nodes directly participating in feedback loops remain susceptible to charge collection events, particularly when the gate is in its evaluation phase (*sleep* de-asserted).

TABLE I: TRAPEZOID CURRENT MODEL PARAMETERS

| Tech. Nodes | T (ps) (N/P) | $t_{rise}$ (N/P) | $t_{fall}$ (N/P) | $pw$ (N/P) |
|---|---|---|---|---|
| 350nm | 95/65 | 20/18 | 280/195 | 130/80 |
| 130nm | 39/28 | 9/6 | 119/80 | 52/42 |
| 45nm | 13/9 | 3/3 | 33/26 | 22/12 |
| **32nm** | **9/6** | **2/2** | **26/19** | **13/9** |

N/P= NMOS/ PMOS

To systematically identify critical sensitivity points, a controlled fault-injection methodology was employed. Trapezoidal current pulses were injected at internal nodes to emulate particle-induced charge deposition, following approaches commonly used in radiation-tolerant asynchronous circuit analysis [8-10]. For each node, the minimum injected charge required to produce a static output-level transition after complete pulse dissipation was extracted and defined as the critical charge, $Q_{\text{crit}}$.

TABLE II. EXTRACTED CRITICAL CHARGE FOR POSITIVE AND NEGATIVE STATIC UPSETS IN A STATIC PCHB NAND GATE.

| Node | $Q_{crit}$ (1→0) | $Q_{crit}$ (0→1) | Worst-Case |
|---|---|---|---|
| Feedback node $\overline{Z^0}$ | ~0.9 fC | ~1.9 fC | **~0.9 fC** |
| Feedback node $\overline{Z^1}$ | ~1.6 fC | ~2 fC | **~1.6 fC** |

Initial evaluation was performed on a static PCHB NAND gate and a static PCHB full-adder cell, the latter representing a deeper arithmetic structure with non-trivial feedback interactions. Both analytical and simulation results indicate that the feedback nodes $\overline{Z^0}$ and $\overline{Z^1}$ exhibit the lowest critical charge thresholds and therefore represent dominant sensitivity points (highlighted in Fig. 1b). These nodes reside within the regenerative hold path and perturbations at these nodes modify the equilibrium of the output rails and therefore, require significantly lower injected charge to induce a static upset.

Table II presents the evaluated $Q_{crit}$ for the sensitive feedback nodes within a PCHB NAND cell. Owing to device strength asymmetry, the critical charge associated with 0→1 and 1→0 static upsets is not necessarily identical. Accordingly, both polarities were independently evaluated, and the minimum value is reported as the worst-case sensitivity metric. For completeness, other internal and output nodes were also evaluated and were found to exhibit substantially higher $Q_{crit}$ values than the feedback nodes. Reported $Q_{\text{crit}}$ values correspond to a load capacitance of 2 fF, considering minimum fanout, under nominal PVT conditions. The trend was consistent for higher load capacitance as well.

Figs. 2b and 2c present representative dynamic and static fault scenarios, respectively. With *sleep* de-asserted and inputs $X = Y =$ DATA1, the gate evaluates to DATA0, corresponding to $Z^1Z^0 = 01$. A transient current pulse is injected at node $\overline{Z^1}$ at 1.25 ns. Fig. 2b illustrates a sub-critical injection event in which the deposited charge ($\approx$ 1.51 fC) is insufficient to exceed $Q_{crit}$. Although a transient perturbation is observed, the regenerative feedback restores the node to its stable operating point, resulting in a dynamically masked glitch without persistent error. In contrast, Fig. 2c depicts a critical injection event in which the deposited charge ($\approx$ 1.65 fC) exceeds the critical threshold. In this case, rather than being attenuated, the disturbance is amplified and latched, forcing the output into the illegal dual-rail state DATAX ($Z^1Z^0 = 11$). Because *sleep* remains de-asserted, regenerative feedback sustains this illegal condition. The persistence of DATAX violates dual-rail mutual exclusion and disrupts QDI protocol, thereby enabling propagation of protocol-level errors to subsequent stages.

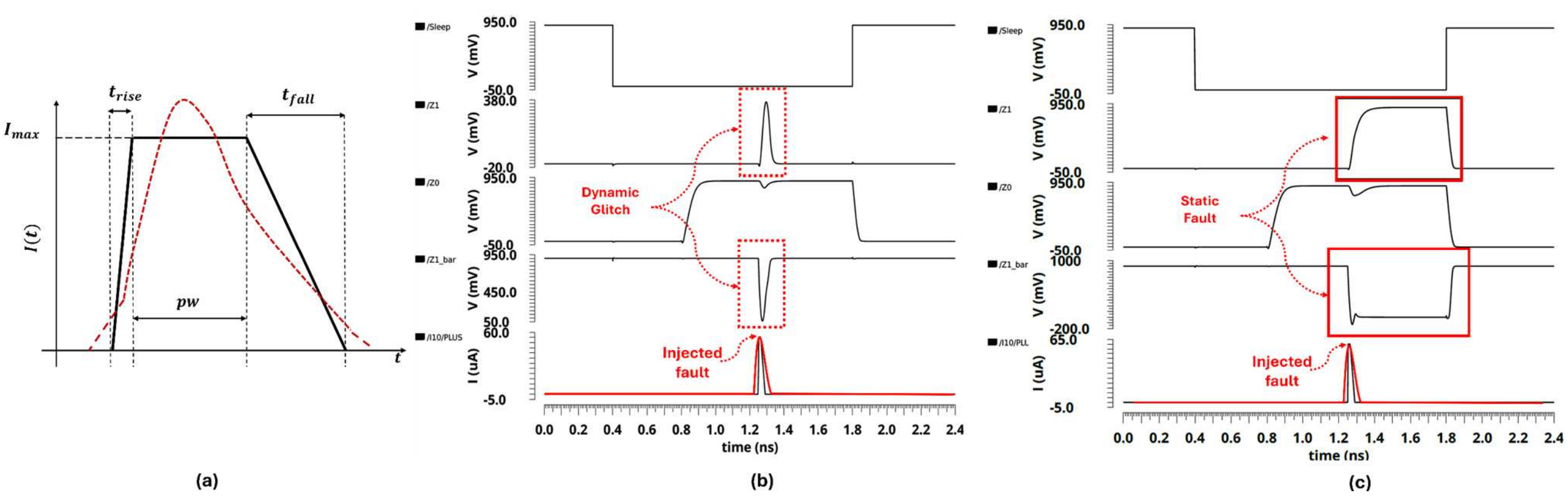

Fig. 2: Error modeling and vulnerability analysis: (a) trapezoid current model, (b) dynamic glitch scenario; and (c) static upset scenario.

## IV. Radiation-Tolerant PCHB Circuit Design

The vulnerability analysis presented in the prior section indicates that static PCHB soft-error resilience is primarily governed by the critical charge of internal feedback-control nodes. Therefore, reliability enhancement efforts must prioritize strengthening the nodes $\overline{Z^0}$ and $\overline{Z^1}$, either by increasing critical charge thresholds or by attenuating transient amplification within the regenerative path. Guided by these findings, subsequent mitigation strategies focus on improving robustness of these nodes to reduce upset probability and prevent fault propagation in larger QDI systems.

### A. Schmitt Trigger-Based Mitigation Approach

Schmitt trigger (ST) structures have been widely employed to suppress transient disturbances in combinational logic, including QDI NCL circuits. In this work, we extend this concept to static PCHB implementations. Schmitt triggers can be categorized as double-sided (DS-ST) or single-sided (SS-ST). A DS-ST provides filtering for both positive and negative transients, whereas an SS-ST selectively suppresses disturbances of one polarity. Prior work on NCL circuits [8] employed SS-ST structures based on the observation that only positive transient glitches and static positive upsets significantly affect functional correctness. Although both NCL and static PCHB incorporate hysteresis, their state-retention mechanisms differ fundamentally. In NCL, output stability during the DATA phase is enforced by threshold-based input consensus. As long as the required number of input rails remain asserted, the pull-up network actively reinforces the output, thereby suppressing negative (1→0) disturbances.

In contrast, static PCHB gates rely on internal regenerative feedback for state holding once evaluation is complete. During the evaluation phase (*sleep* de-asserted), the output DATA must be maintained until handshake completion. A negative static upset occurring in this phase can induce a premature NULL transition prior to sleep assertion. Because the PCHB protocol requires DATA retention until acknowledgment, such a 1→0 event constitutes a protocol violation. The downstream stage

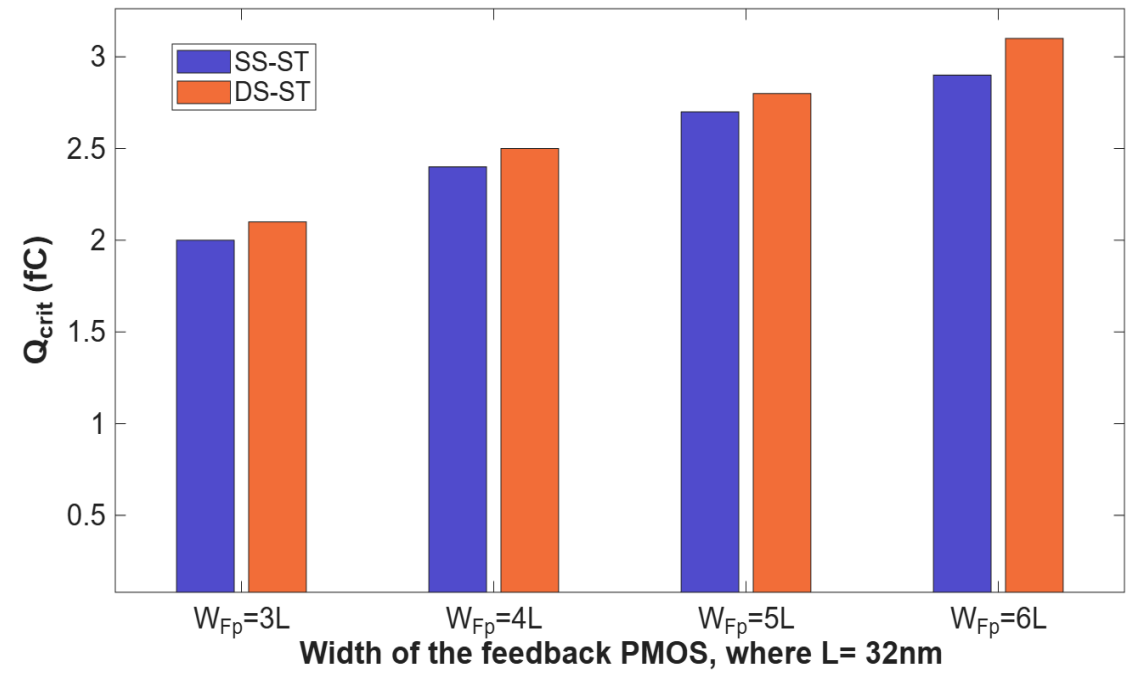

Fig. 4. Effect of feedback transistor sizing on $Q_{crit}$ for static PCHB NAND.

may lose its DATA token before completing evaluation, potentially leading to deadlock.

Thus, unlike NCL, static PCHB circuits exhibit polarity-symmetric vulnerability and both 0→1 and 1→0 static upsets are reliability-relevant. As shown in Fig. 2c, a positive static upset may drive the output into the illegal dual-rail state DATAX (11), while a negative static upset during evaluation may prematurely force a transition to NULL (00), disrupting QDI handshake correctness. Given that both polarities can induce protocol-level failures in static PCHB, we first adopt a DS-ST approach to filter disturbances in both directions. For comparison, we also adopt the SS-ST approach, which provides lesser and asymmetric protection, and allows evaluation of the trade-off between resilience and design overhead (Section V).

The sizing of the feedback transistors, *Fp* and *Fn* (Fig. 3), directly influences the soft-error resilience of both the DS-ST and SS-ST configurations. To quantify this effect, Fig. 4 presents a parametric sweep of the feedback PMOS width for a static PCHB NAND, varied from 96 nm (3×L) to 192 nm (6×L). The results show a clear increase in the critical charge $Q_{\mathrm{crit}}$ at node $\overline{Z^1}$ as the feedback PMOS width increases, indicating improved robustness with stronger feedback. The enhancement is slightly more pronounced in the DS-ST configuration compared to SS-ST. A similar trend was observed in the full-adder implementation as well.

### B. Transmission Gate Based Mitigation Approach

To mitigate the dominant sensitivity of the internal feedback-control nodes $\overline{Z^0}$ and $\overline{Z^1}$, a transmission gate (TG)-based attenuation strategy is introduced, as illustrated in Fig.

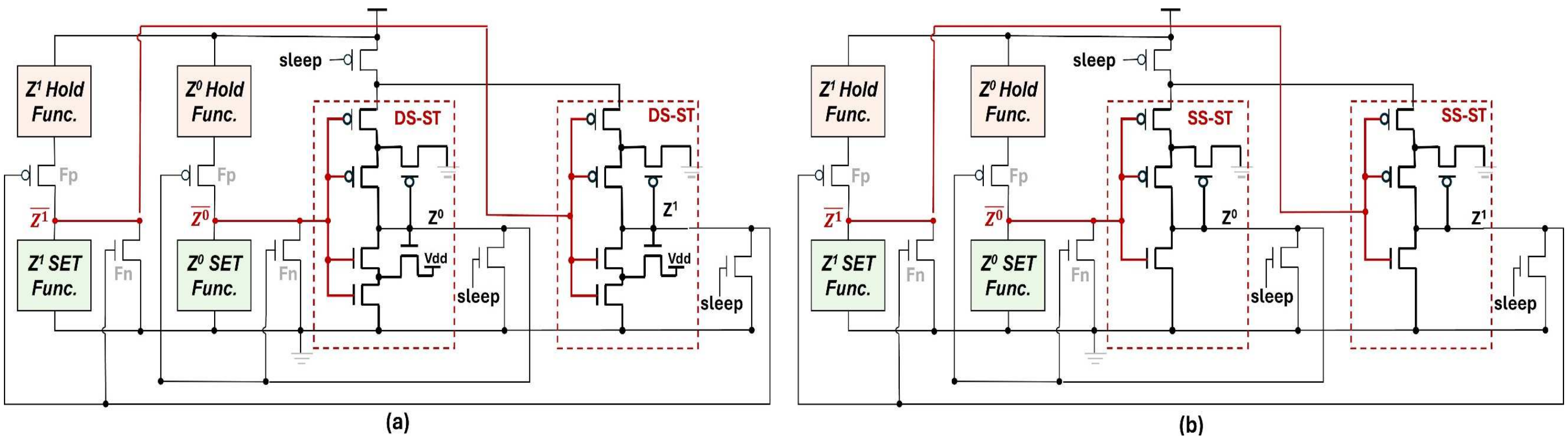

Fig. 3: (a) Static PCHB with DS-ST template and (b) Static PCHB with SS-ST template. Note: Fp and Fn are feedback P/NMOS, respectively.

5a. The proposed structure inserts a TG between the internal regenerative node and the corresponding output rail ($Z^0, Z^1$), thereby electrically decoupling the feedback path from high-frequency transient perturbations. The TG consists of parallel NMOS and PMOS devices biased to remain continuously ON. Unlike a conventional logic-controlled pass element, the TG in this configuration functions as a controlled resistive element rather than a switch. By appropriately selecting gate bias voltages, $V_N$ and $V_P$, TG introduces a finite series resistance ($R_{TG}$) while preserving full logic-level transmission under steady-state conditions. The ON-state conditions satisfy $V_{GS,N} \geq V_{TH,N}$ for NMOS and $V_{SG,P} \geq |V_{TH,P}|$), where $V_{GS,N}$ and $V_{SG,P}$ are the gate-to-source/source-to-gate voltages of NMOS/PMOS, respectively, and $V_{TH,N}$ and $V_{TH,P}$ are the threshold of NMOS and PMOS, respectively. This ensures that valid DATA transitions propagate without logic-level degradation.

The mitigation effect arises from two coupled phenomena: current attenuation and RC filtering. During particle-induced current surge at $\overline{Z^0}$ or $\overline{Z^1}$, the regenerative feedback attempts to translate the disturbance into the output rail ($Z^0$ or $Z^1$). The series resistance of the TG, $R_{TG}$, limits the instantaneous current that can flow into the rail node. In addition, $R_{TG}$, combined with the effective capacitance of the output node, forms a low-pass RC filtering network. Short-duration glitches are attenuated before the regenerative loop can amplify them. Consequently, the effective perturbation magnitude at the output node is reduced. As a result, a larger injected charge is required to affect the stability of the internal state of the circuit, thereby increasing $Q_{crit}$.

The effectiveness of the TG-based attenuation depends on the chosen bias voltages $V_N$and $V_P$, which directly control the ON resistance. Fig. 5b compares three $V_N/V_P$ bias configurations: 0.9/0.0 V (full enhancement), 0.8/0.1 V, and 0.7/0.2 V. As $V_{GS,N}$ and $V_{SG,P}$ are reduced, the TG resistance increases, leading to stronger attenuation of transient disturbances. This is evident from the reduced glitch amplitude in Fig. 5b. However, excessive resistance may degrade propagation delay and increase dynamic energy due to slower rail transitions [10]. Therefore, the TG bias must be selected to balance resilience improvement against performance overhead.

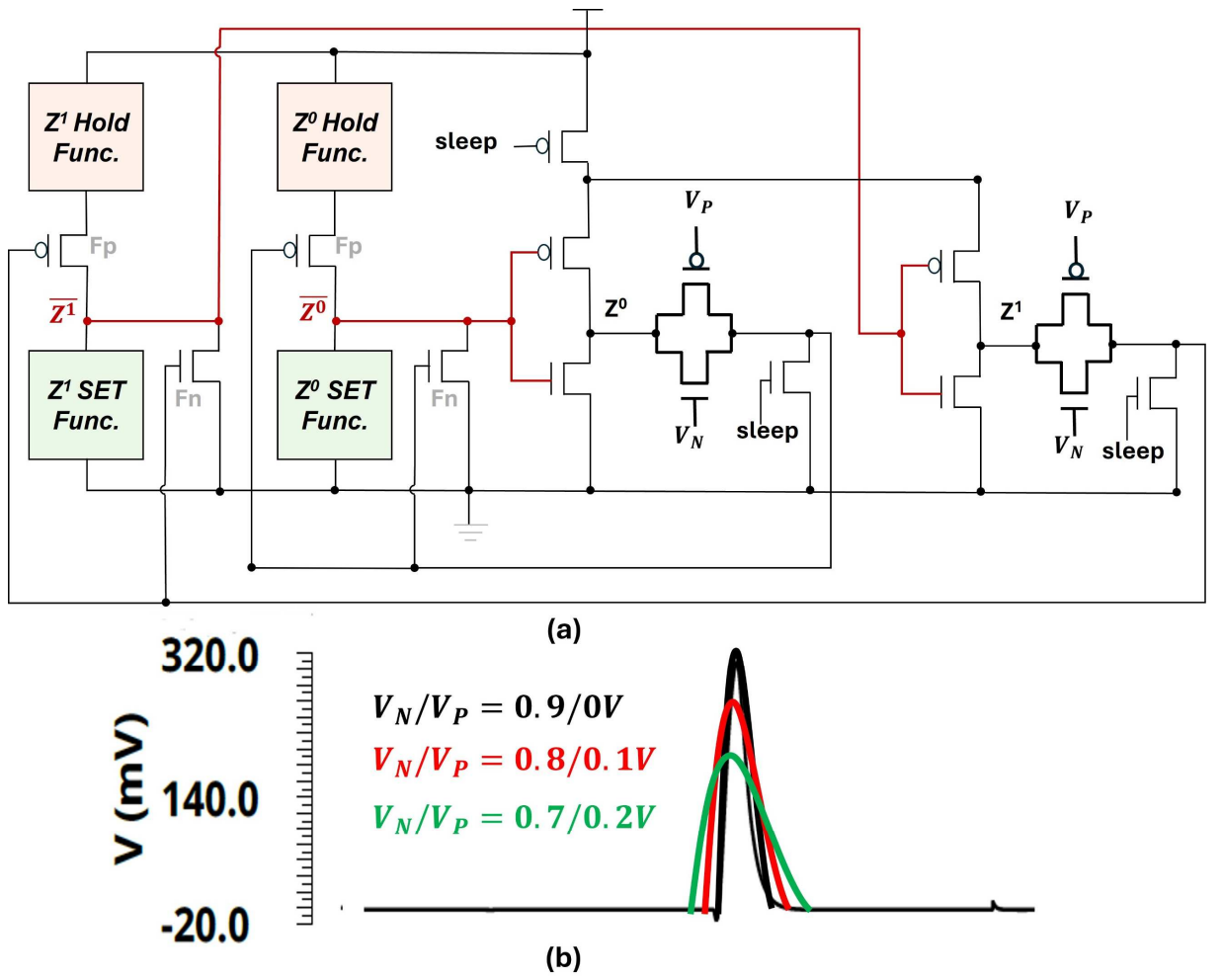

Fig. 5. a) Static PCHB with transmission gates, and b) effect of TG bias configuration on glitch attenuation.

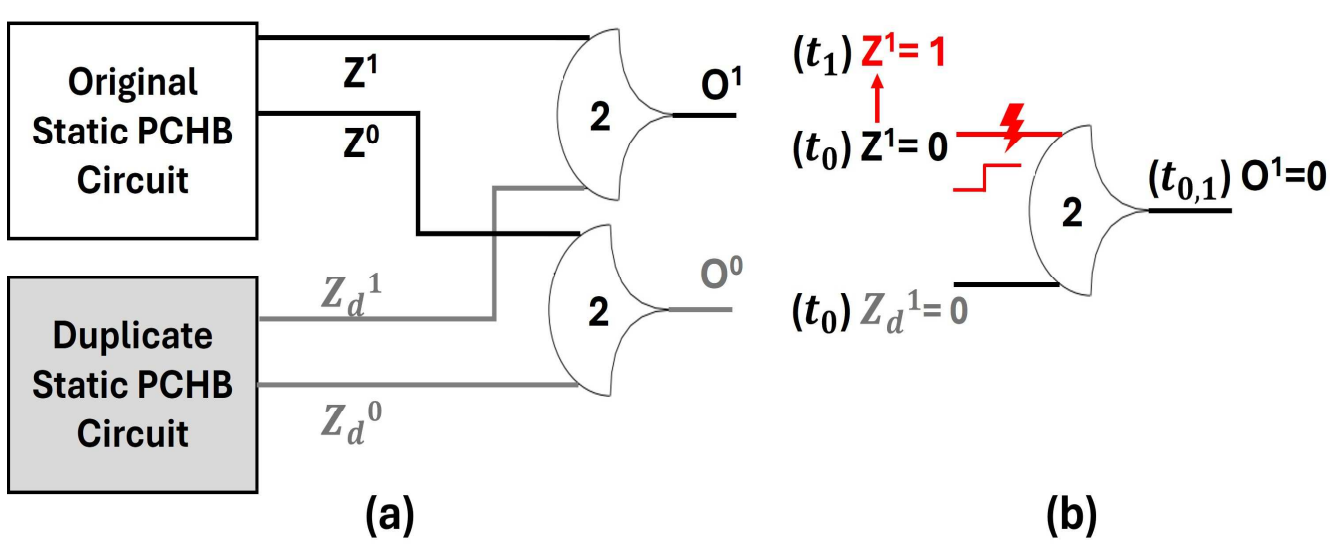

Fig. 6. a) Duplication based PCHB design template, and b) representative fault scenario.

### C. Duplication Based Mitigation Approach

The duplication-based mitigation strategy follows Jang and Martin's redundancy principle [12], wherein the original static PCHB circuit is replicated to form two structurally identical copies, as depicted in Fig. 6(a). Each rail of a dual-rail output in the primary circuit is combined with its corresponding rail from the duplicated instance using a 2-input C-element. The C-element output constitutes the final circuit output. Functionally, the C-element implements consensus-based filtering. It updates its output only when both inputs agree, i.e., both are asserted or both are de-asserted. If the two inputs differ, the C-element holds its previous state. This property enables temporal and spatial masking of single-event-induced static upsets occurring in only one replica. Unlike the ST and TG-based approaches, which focus on enhancing the robustness of sensitive internal nodes, the duplication strategy does not modify the intrinsic vulnerability of each individual copy. Instead, it introduces architectural redundancy, relying on output-level consensus to prevent fault propagation. The effectiveness depends on the assumption that the probability of simultaneous correlated faults in both replicas is negligible.

Fig. 6(b) illustrates a representative fault scenario. Under nominal conditions at time $t_0$, both the $Z^1$ rail of the original circuit and the corresponding $Z_d^1$ rail of the duplicated circuit are at logic '0'. Consequently, the TH22 gate (functionally equivalent to a 2-input C-element) outputs '0'. At time $t_1$, a particle strike in the original circuit induces a positive static upset, causing $Z^1$ to transition to '1'. However, since the duplicated circuit remains fault-free and continues to output '0', the C-element maintains its previous state, thereby blocking propagation of the erroneous transition to the final output $O^1$.

## V. Simulation Results and Discussions

Five representative static PCHB standard cells, NAND, NOR, XOR, Half Adder (HA), and Full Adder (FA), were implemented under five configurations: baseline (unhardened), DS-ST, SS-ST, TG, and duplication-based redundancy. In total, 25 distinct circuit implementations were designed and evaluated. All simulations were performed at the transistor level using Cadence Spectre with the 0.9 V 32-nm high-performance CMOS predictive technology model (PTM) [13]. A load capacitance of 2 fF was selected to approximate a minimum fanout condition at 32 nm. Since reduced load capacitance lowers $Q_{\text{crit}}$ and maximizes voltage perturbation for a given deposited charge, this choice represents a conservative worst-case sensitivity condition for evaluating intrinsic hardening effectiveness.

The designs were compared using three key metrics: propagation delay ($T_{DD}$), switching energy per operation, and transistor count. $T_{DD}$ denotes the full token-cycle latency, defined as the sum of the DATA evaluation interval, sleep-generation interval, and return-to-NULL interval. Switching energy is computed for one complete cycle of execution. For delay and energy, all valid input combinations were simulated, and reported values correspond to the arithmetic mean across cases to ensure unbiased comparison. To maintain fairness, transistor sizes were kept identical across baseline and hardened implementations. Device dimensions were selected to guarantee correct functional operation under nominal PVT conditions for all designs. Hardening techniques were applied exclusively to the PCHB logic and registration structures. Sleep-generation circuitry was intentionally left unmodified to isolate the impact of hardening within the core PCHB datapath. Since sleep-generation blocks comprise NCL gates, existing NCL-specific hardening methods could be independently applied if desired.

Simulation results are summarized in Tables III–V, while Table VI reports the average normalized overhead in propagation delay, switching energy, and transistor count across all evaluated cells. Among the mitigation strategies, SS-ST and TG introduce the smallest transistor-count overhead, averaging 6.1% relative to the baseline implementations. However, as discussed in Section IV-A, SS-ST selectively suppresses only one polarity of transient disturbance and therefore does not provide comprehensive protection for static PCHB circuits. Consequently, despite its low structural overhead, SS-ST is not an appropriate standalone hardening solution for static PCHB. Duplication-based redundancy incurs a substantially higher area cost, with an average transistor-count increase of 81.6% over baseline designs. If the sleep-generation units were also replicated, the overall area overhead would exceed 100%, underscoring the structural expense associated with architectural redundancy. With respect to performance metrics, TG-based implementations exhibit the lowest propagation delay overhead among the hardened configurations. DS-ST, while providing strong polarity-symmetric robustness, incurs the largest delay penalty. A similar trend is observed in switching energy: TG achieves the most favorable energy profile, whereas DS-ST utilizes the highest energy per operation. Overall, these results highlight a clear resilience-overhead trade-off.

**TABLE III.** PROPAGATION DELAY, TDD (PICOSECONDS)

| Circuit | Baseline | SS-ST | DS-ST | TG | Duplicate |
|---|---|---|---|---|---|
| NAND | 128.6 | 176.2 | 182.2 | 157.2 | 189.7 |
| NOR | 128.2 | 167.8 | 181.3 | 156 | 183.1 |
| XOR | 139.5 | 221.2 | 233.3 | 168.7 | 191.1 |
| HA | 152.8 | 224.5 | 226.7 | 182.5 | 205.7 |
| FA | 182.9 | 247.8 | 254.3 | 208.5 | 238.5 |

**TABLE IV.** SWITCHING ENERGY PER OPERATION (FJ)

| Circuit | Baseline | SS-ST | DS-ST | TG | Duplicate |
|---|---|---|---|---|---|
| NAND | 3.1 | 5.3 | 5.6 | 3.5 | 4.7 |
| NOR | 3.1 | 5.3 | 6.1 | 4.2 | 4.5 |
| XOR | 3.3 | 6.9 | 7.4 | 4.0 | 4.9 |
| HA | 6.3 | 10.5 | 11.3 | 7.4 | 9.4 |
| FA | 6.0 | 9.6 | 13.7 | 8.3 | 10.4 |

**TABLE V.** NUMBER OF TRANSISTORS

| Circuit | Baseline | SS-ST | DS-ST | TG | Duplicate |
|---|---|---|---|---|---|
| NAND | 63 | 67 | 71 | 67 | 106 |
| NOR | 63 | 67 | 71 | 67 | 106 |
| XOR | 71 | 75 | 79 | 75 | 122 |
| HA | 98 | 106 | 114 | 106 | 192 |
| FA | 162 | 170 | 178 | 170 | 304 |

**TABLE VI.** AVERAGE NORMALIZED OVERHEAD AS COMPARED TO THE UNHARDENED IMPLEMENTATION

| Circuit | Delay | Energy | Transistor Count |
|---|---|---|---|
| **SS-ST** | +41.7% | +72.4% | **+6.1%** |
| **DS-ST** | +47.2% | +102% | +12.2% |
| **TG** | **+19.2%** | **+25.7%** | **+6.1%** |
| **Duplication** | +37.7% | +55% | +81.6% |

## VI. CONCLUSIONS AND FUTURE WORK

This work presents a systematic soft-error characterization and hardening trade-off analysis of static PCHB circuits. Unlike threshold-consensus-based NCL gates, static PCHB circuits exhibit unique transient fault behavior due to their regenerative feedback-based state-retention mechanism. Transistor-level fault injection identifies internal feedback-control nodes as the most vulnerable points, with the lowest critical charge. Four mitigation techniques, DS-ST, SS-ST, TG reinforcement, and duplication, are evaluated across representative PCHB standard cells, revealing clear resilience–overhead trade-offs. Future work includes layout-aware modeling and system-level evaluation of hardened static PCHB architectures.